\documentclass[psfig,fleqn,twoside]{article}
\usepackage{espcrc2}

\usepackage{graphicx}
\usepackage[figuresright]{rotating}

\newcommand{\AmS}{{\protect\the\textfont2
  A\kern-.1667em\lower.5ex\hbox{M}\kern-.125emS}}

\title{Search for MSSM Higgses at the Tevatron}

\author{A. Connolly 
\address[MCSD]{Lawerence Berkeley National Laboratory, 50B-5239, One Cyclotron Rd., Berkeley, CA  94720  U.S.A.
	}%
        \thanks{Current address is Fermi National Accelerator Laboratory, P.O. Box 500, M.S. 318, Batavia, IL  60510  U.S.A.}
\\For the CDF and D0 Collaborations 
      }
\begin{document}

\begin{abstract}
We present an overview of searches for MSSM Higgs at the Tevatron, 
concentrating on searches probing the high $\tan{\beta}$ region.  We
discuss the search for $A/H \rightarrow \tau \tau$ which is soon to 
be completed in the Run I data and review the new tau triggers 
implemented by CDF and D0 in Run II, which will greatly impact this 
analysis. 
We also
present the results of a Run I search for $A/H bb \rightarrow bbbb$ performed
by CDF and highlight expected improvements in this channel by both experiments
in Run II.

\vspace{1pc}
\end{abstract}

\maketitle

\section{MOTIVATION}

The Higgs mechanism breaks electroweak symmetry in the Standard Model,
giving mass to particles through its couplings.  Current data from 
electroweak precision measurements points to a light Higgs 
($M_{Higgs} < 199$ GeV @ 95\% CL \cite{PDG}).  However, the Higgs has never been
definitively observed ($M_{Higgs}>114$ GeV at 95\% CL \cite{LEP2}).

A Standard Model Higgs suffers from 
the so-called hierarchy problem.  The theory needs 
fine-tuned parameters to accomodate
a light Higgs mass.  Supersymmetry offers
a solution to this problem, through a symmetry between fermions and bosons.  

The
Minimal Supersymmetric Standard Model (MSSM) contains two Higgs doublets,
leading to five physical Higgs bosons:  Two neutral 
CP-even states (h and H), one neutral CP-odd (A), and two charged states
($H^+$ and $H^-$).  At tree-level,
the masses are governed by two parameters, often taken to 
be $m_A$ and $\tan{\beta}$ \cite{MSSM}.    
When $\tan{\beta} >> 1$ , A is nearly degenerate with one of the
CP-even states (denoted $\phi$).  Where $m_A \leq 130$ GeV ($m_A \geq 130$),  
$m_A \cong m_h$ ($m_A \cong m_H$).  

In this same large $\tan{\beta}$ region, the cross sections for some
production mechanisms such as $pp \rightarrow A(\phi)$ and
$pp \rightarrow A(\phi) b\bar{b}$ are enhanced  by factors of 
$\tan{\beta}^2(\sec{\beta}^2)$.  For example,
with $\sqrt{s}=2$ TeV, $\tan{\beta}=30$ and $m_A=100$ GeV, the
cross sections for $pp \rightarrow A$ and $pp \rightarrow \phi$ are 
each of order 10 pb\cite{Spira}.  The cross section for $pp \rightarrow A/\phi b\bar{b}$
is smaller, but within the same order of magnitude.
In the same region, the branching ratios
to $A/\phi \rightarrow b\bar{b}$ and $\tau\tau$ dominate, at
$\sim 90\%$  
and $\sim10\%$ respectively, independent of mass. 

Due to their similar masses, cross-sections and branching ratios in the
 high $\tan{\beta}$ region,
we search for both A and $\phi$ simultaneously.  At the Tevatron, we search
for $pp \rightarrow A/\phi \rightarrow \tau \tau$ 
(the $b\bar{b}$ final state is expected
to be overwhelmed by dijet background) and $pp \rightarrow A/\phi b\bar{b}
\rightarrow b\bar{b}b\bar{b}$.

\section{SEARCH FOR $pp \rightarrow A/\phi \rightarrow \tau^+ \tau^-$}

This search is underway at CDF. The dominant issues for this analysis are:  tau identification,
ditau mass reconstruction, irreducible background from $Z \rightarrow \tau \tau$, and event loss at the trigger level.

Wherever not specified, we use the benchmark case of 
$m_A=95$ GeV and $\tan{\beta}=40$ to quote efficiencies and cross-sections.

\subsection{Tau Identification}

Compared to QCD jets, taus are highly collimated, leaving narrow jets with
 low track and photon multiplicity, and low mass.

In CDF, when selecting taus, one typically requires a jet with high
visible $E_T$ containing a high $p_T$ track.  The jet is required to 
be isolated in a $10^o-30^o$ annulus around the high $p_T$ track.
The visible energy in a $10^o$ cone is required to   
satisfy low track and photon 
multiplicity requirements and to reconstruct a mass $m<1.8$ GeV.  
Additionally, a requirement
is made on the charge of the tracks in the $10^o$ cone when appropriate.
In Run I,  CDF acheived fake rates in the range 1.2 - 0.7 \% for jet $E_T$
between 20 and 200 GeV\cite{tau_fakes}.

\subsection{Ditau Mass Reconstruction}

The full mass of a ditau system may be reconstructed \cite{CMS}
if the neutrinos are assumed to travel in the same direction
as their parent taus, by solving the following system of equations:

\begin{equation}
\label{eq:mass1}
E_x^{meas}\hspace{-0.35in}/\hspace{0.30in} = E_x^{\tau 1}\hspace{-0.20in}/\hspace{0.13in} + E_x^{\tau 2} \hspace{-0.20in}/\hspace{0.13in}
\end{equation}
\begin{equation}
\label{eq:mass2}
E_y^{meas}\hspace{-0.35in}/\hspace{0.30in} = E_y^{\tau 1}\hspace{-0.20in}/\hspace{0.13in} + E_y^{\tau 2}\hspace{-0.20in}/\hspace{0.13in}
\end{equation}
where $E_{x,y}^{meas}\hspace{-0.35in}/\hspace{0.30in}$ are the x and y components of the measured event missing energy, and $E_{x,y}^{\tau 1}\hspace{-0.24in}/\hspace{0.20in}$ 
and $E_{x,y}^{\tau 2}\hspace{-0.24in}/\hspace{0.20in}$ denote the missing energy from each tau.

Equations \ref{eq:mass1} and \ref{eq:mass2} do not give a meaningful solution when the taus are back-to-back in the
transverse plane.  Therefore, we require that 
$|\sin\Delta\phi|>0.3$, where $\Delta\phi$ is the azimuthal angle between
the tau candidates.

When the solution to Equations \ref{eq:mass1} and \ref{eq:mass2} gives  
$E^{\tau 1}\hspace{-0.20in}/\hspace{0.18in}<0$ 
or $E^{\tau 2}\hspace{-0.20in}/\hspace{0.18in}<0$, the event is thrown
out, causing about 50\% of the Higgs signal to be lost.  However, 97\%
of W+jets events are rejected in this way, which would otherwise be
a formidable background.

We generate $A/\phi \rightarrow \tau \tau$ events in Pythia 6.203 with $m_A=95$ GeV and $\tan{\beta}=40$.
After simulation of the Run I CDF detector, a ditau mass
distribution is reconstructed with a mean value of 93.7 GeV with an RMS of 24.1 GeV.

\subsection{Irreducible Background}

The dominant \emph{reducible} backgrounds to this analysis are 
QCD, $Z \rightarrow e e$, and W+jets.   
$Z \rightarrow \tau \tau$ is an irreducible background, but
Higgs events are more
efficient for this search than Z's for a couple of reasons.

First, in the high $\tan{\beta}$ region, $A/\phi$'s have a high branching
ratio to taus (9\%) compared to Z's (3.7\%).  Second, an $A/\phi$ is typically produced with a stiffer $p_T$
than a Z.  This means that the requirement $|\sin\Delta\phi|>0.3$, which is
nearly equivalent to $p_T^{A/\phi/Z}>15$ GeV, is $\sim 30\%$ more efficient for Higgs events than Z events.

\subsection{Triggers}

Since there was no $\tau$ trigger in Run I at CDF, the analysis uses
a lepton trigger requiring $p_T>18$ GeV, seeking events with one
leptonic and one hadronic decay.  Since only half of the
signal events decay in this way, and of these, only 20\% 
contain a lepton which satisfies the $p_T$ requirement within 
the acceptance, the signal rate is greatly diminished
at the trigger level.

This major loss at the trigger level is problematic,
since the cross section drops
by a factor of 4 from $m_A=95$ GeV to $m_A=120$ GeV, before the
mass reconstruction, with an RMS of 24 GeV, can discriminate from  
$Z \rightarrow \tau \tau$.  Therefore,
in Run II, CDF and D0 are both implementing triggers designed for tau
physics.  Lowered $p_T$ thresholds and 
new decay modes available 
will greatly increase the acceptance for this search.

In Run II, CDF and D0 both have lepton + track triggers and $\tau+E_T\hspace{-0.18in}/\hspace{0.1in}$ triggers.  In addition, both experiments are triggering
on events with two hadronic taus.  CDF's trigger is calorimeter-based,
while D0's is track-based.  

The Run I search for $A/\phi \rightarrow \tau \tau$ is still work in 
progress, and the Run II analysis is also in the works.

\section{SEARCH FOR $pp \rightarrow A/\phi b\bar{b} \rightarrow b\bar{b}b\bar{b}$}

CDF performed this search in Run I.  Both experiments expect to 
improve on the analysis in Run II. 

\subsection{Run I search}

The Run I search at CDF \cite{bbbb}utilized a 4-jet trigger requiring $\Sigma E_T>125$ GeV.  Three b-tags were required based on
displaced vertices, and the b jets were required
to be separated in azimuthal angle, $\Delta\phi>1.9$.  To 
optimize sensitivity, the 
$E_T$ cuts on the jets  varied
with mass hypothesis.  For mass hypothesis
below 120 GeV (above 120 GeV), the second and third b-tagged jets (first and second jets) ordered in $E_T$
were chosen for the mass reconstruction.  The search is performed in mass windows dependedent on mass hypothesis.

The product of branching ratio and acceptance ranged from 0.2 to 0.6\%
in the mass range 70 and 300 GeV.  For a mass hypothesis of 70 GeV,
5 events were observed with 4.6 $\pm$ 1.4 expected.  Only these 5 events
appear in the higher mass windows.  No excess above predicted is observed.
Figure 1 shows the $m_A-\tan{\beta}$ region excluded.

\begin{figure}[htb]
\vspace{9pt}
\includegraphics*[height=2.0in]{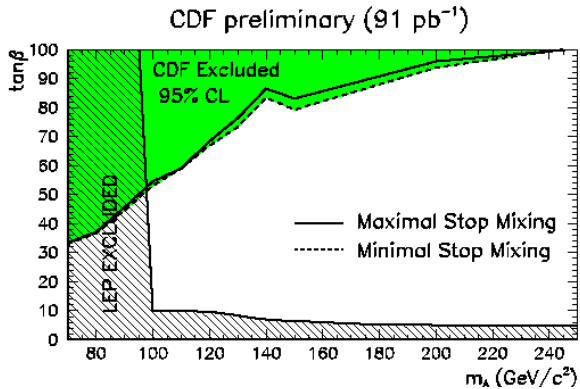}
\caption{Region of the $m_A-\tan{\beta}$ region exluded by the CDF search.}
\label{fig:bbbb_exclusion}
\end{figure}

\subsection{Run II Improvements}

At CDF, studies of $Z \rightarrow b\bar{b}$ events show an improved
dijet mass resolution after correcting for muons, 
$E_T\hspace{-0.18in}/\hspace{0.1in}$, and nonlinearities in the
 hadronic calorimeter.  Separate studies of
QCD jets using similar techniques show a 30\% improvement in jet resolution.

B-tagging in Run II at CDF will be improved with the new ability to reconstruct
three-dimensional tracks.  Extended coverage from $|\eta|<1$ (Run I) to $|\eta|<2$ means improve b-jet and lepton acceptance.  Additionally, new triggers will
also recover acceptance, including a displaced track trigger, and an improved multijet trigger.

With a new silicon detector, D0 will also be performing this search in
Run II, expecting a 12\% dijet mass resolution.  Both experiments
perform a study of their expected sensitivity to   $pp \rightarrow A/\phi b\bar{b} \rightarrow b\bar{b}b\bar{b}$ in Run II, and obtain
similar results\cite{run2higgs}.  We present the D0 study here.

D0 also uses a multijet trigger, requiring four jets, each with $E_T>15$ GeV.
To maximize sensitivity, mass dependent $E_T$ cuts are made on the jets.  
At least 3 b tags are required.  
All mass combinations are plotted,
and a 2.5$\sigma$ $b\bar{b}$ mass window is used.
With 2$fb^{-1}$, D0 concludes that the Tevatron 
is expected to exclude $m_A<160$ GeV for $\tan{\beta}=40$ at 95\% CL, and
a 5$\sigma$ discovery for $m_A<115$ GeV for the same $\tan{\beta}$.

\section{CONCLUSIONS}

Run I results of the search for $A/\phi \rightarrow \tau \tau$ at CDF  
are to be completed soon, and a first glimpse of Run II data is on the
way.

The Run I search for $pp \rightarrow A/\phi b\bar{b} \rightarrow b\bar{b}b\bar{b}$ derives lower mass limits for $\tan{\beta}$ in excess of 35.  
In Run II with
both experiments searching for this decay mode, the Tevatron
is expected to exclude (or make a discovery in) a significant region
of MSSM parameter space.  Both experiments are optimistic about
improvements from triggers, jet resolution, and b-tagging to make
this search even stronger than the current projections.


\begin{thebibliography}{9}
\bibitem{PDG} K. Hagiwara et al., Physical Review D66, 010001 (2002).
\bibitem{LEP2} ALEPH Collaboration, DELPHI Collaboration, L3 Collaboration,
		OPAL Collaboration and LEP Higgs Working Group, 
		LHWG Note/2001-03,hep-ex/0107029.
\bibitem{MSSM} For a review of the MSSM, see H.P. Nilles, Phys. Rept. 110:1,1984, H. Haber, G. Kane, Phys. Rept. 117:75,1985.
\bibitem{Spira} M. Spira, Nucl.Instrum.Meth.A389:357-360,1997
\bibitem{tau_fakes} CDF Collaboration, Phys.Rev.Lett.79:357-362, 1997
\bibitem{CMS} CMS Technical Proposal, CERN/LHCC 94-38, December 1994, pp. 191-192. 
\bibitem{bbbb} CDF Collaboration, Phys.Rev.Lett.86(2001) 4472-4478.
\bibitem{run2higgs}M.Carena, J.S.Conway, H.E.Haber, J.D.Hobbs, et al, Fermilab-Conf-00/279-T and SCIPP-00/37,hep-ph/0010338.
\end{thebibliography}
\end{document}